\documentclass{article}

\usepackage[dvips]{graphicx}
\usepackage{epsfig}
\usepackage{longtable}
\title{Acoustic Spectroscopy of the DNA in GHz range}
\author{V.N.Blinov\footnote{blinov.veniamin@gmail.com}   and V.L.Golo\footnote{voislav.golo@gmail.com}\\
             Department of mechanics and mathematics \\
             the Lomonosov Moscow State University\\ Moscow, Russia}

\begin{document}

\maketitle

\begin{abstract}
We find a parametric resonance in the GHz range of the DNA dynamics, generated by pumping hypersound .
There are localized phonon modes caused by the random structure of elastic modulii due to the sequence
of base pairs.
\end{abstract}

\section{Introduction: phonon modes of the DNA}

Molecule  of the DNA has unusual elastic properties owing to its helicoidal symmetry and   the sequence of base-pairs.
From the physical view-point, i.e. neglecting its genetic information, the latter looks   random. The vibrational dynamics of  the DNA
has an important bearing upon biological phenomenae in cell, \cite{Mei}, see also\cite{Ju}.  A specific feature of the dynamics is
localized motions in the duplex which spread only over several base-pairs. Their existence has been confirmed by the experimental research
using   the Raman scattering, \cite{Urabe},\cite{Tominaga}, \cite{Rupprecht}, \cite{Edwards}, the far-infrared absorption,
\cite{Prabhu},\cite{Powell}, and the Brillouin scattering \cite{Lindsay}. In paper \cite{Woolard} the vibrational modes have been studied
using the submillimeter-wave absorption spectroscopy in the range of  $ \sim 0.01 - 10 $  THz. Thus, Woolard et al, \cite{Woolard},
have found multiple dielectric resonances in the long-wavelength portion of the submillimeter-wave regime, i.e. $\sim 1 - 30 cm^{-1}$,
which they ascribe to  phonon modes of the DNA. These results   provide valuable methods of bio-diagnostics, \cite{Woolard2}.

Theoretical study of phonon modes of the DNA had revealed  elastic excitations of
the  duplex which may correspond to the approximate helicoidal symmetry of a molecule of the DNA,
and its random elastic structure, \cite{Putnam}, \cite{Saxena_Zandt}.
It is important that
phonon modes of the DNA are believed to be  strongly attenuated owing to an interaction with ambient medium.  The effect can be mitigated, to some extent, by preparing samples  of appropriate character. Thus, films formed by molecules of the DNA appears to be less prone to the attenuation  as regards phonon modes. In contrast, it is  especially strong in liquid solutions of the DNA.
It should be noted that  the estimates  are  based on the Navier-Stokes hydrodynamics in {\it sub}-GHz range, even more so, the classical Stokes formula for the viscous drag at small Reynolds number.  But the above arguments are  wrong in the GHz range.  In fact, Van Zandt, \cite{VanZandt}, showed that within the framework of the Maxwell hydrodynamics, there could exist underdamped phonon modes. Thus,  whether the phonon modes are damped or not so damped,  remains to be seen.

In this  paper we  follow the analysis performed by Chia C.Shih and S.Georghiou who have put forward powerful arguments in favour of
the existence of underdamped vibrational modes of a molecule of the DNA,  besides the overdamped ones \cite{Georghiou}.
The authors visualize a molecule of the DNA as a duplex of two strands formed by sugar-phosphate backbones framed by base-pairs located inside the strands, and  assume that the bases are shielded by sugar---phosphates from the bombardment by molecules of solvent,  the base-pairs being influenced by  the ambient medium indirectly, through their interaction with the backbone. The motion of the bases has the shape of librations inside the cages formed by the sugars of the backbone. Consequently, the dynamics of the backbone is damped owing to the strong attenuation caused by the medium, whereas that of the base-pairs turns out to be underdamped.  Their angular frequencies are insensitive to the viscosity and lie in the low  range of the Raman spectrum.  Thus, the backbone modulates the motion of base-pairs in accord with the environment medium.

We    aim at finding effects of resonance attenuation of hypersound propagating in a sample of molecules of the DNA.
To that end one may employ the interaction between solvent and molecules of the DNA
to generate phonon modes of the DNA by pumping GHz-excitations in the liquid. The strong viscous interactions of the molecules and the solvent
may result in a dragging  that could promote torsional phonon modes   generating the interstrand ones, and
result in the additional absorption of hypersound. Thus,  the study of solutions of the DNA could provide important information as to the nature of the liquid state,
precisely through taking into account  effects of dissipation in solvent in the GHz range.
It is worth noting that  hypersound acoustics has made considerable progress in recent years  and has become  a powerful
method in experimental research, \cite{Huynh}, \cite{Linde}. In the context of the DNA  it may turn out to be a valuable means for studying hydrodynamic
phenomenae.


\section{Librational dynamics of base-pairs}
In this paper we are concerned with the phonon modes of  the DNA.
To that end we employ the theoretical model worked out in our earlier paper \cite{Golo}, in which we follow the guidelines cast
by H.Kappellmann and W.Biem,  \cite{Biem}. For the convenience of the reader we recall the main points of paper \cite{Golo}.
In considering the dynamics of the DNA one has to take into account:
\begin{enumerate}
  \item the DNA having the two strands; \\
  \item the base-pairs being linked by the hydrogen bonds; \\
  \item the helical symmetry.
\end{enumerate}

\begin{figure}[h]
	\center{\includegraphics[width=11cm]{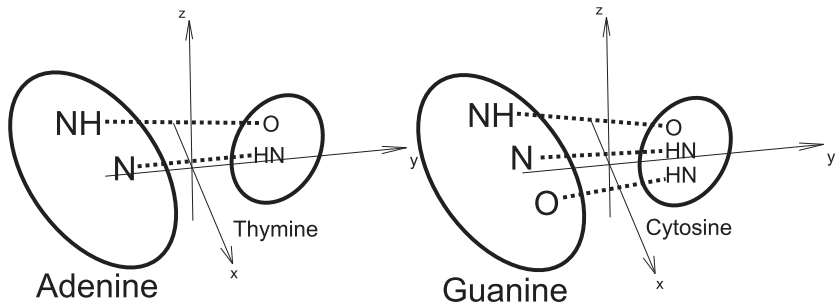}}
	\caption{Base-pairs: adenine---thymine and guanine---cytosine. Dashed lines designate hydrogen bonds.}
	\label{fig:pairs}
\end{figure}

\begin{figure}[h]
	\center{\includegraphics[width=11cm]{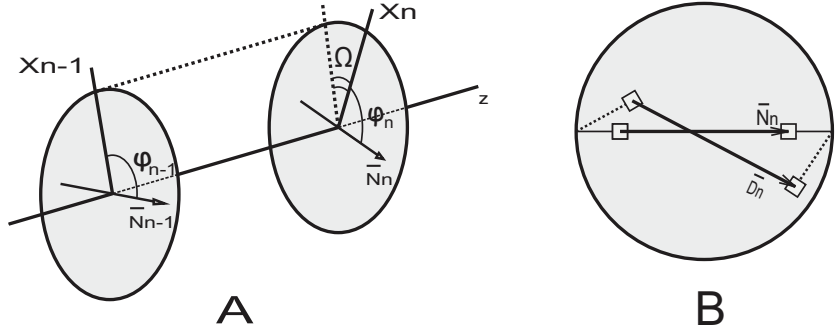}}
	\caption{ Angles $\varphi_n$ indicate deviation from equilibrium. Angle $\Omega = \pi/5$ denotes the pitch of the double helix.
	          Vectors $\vec{N}_n$ and $\vec{D_n}$ designate the unperturbed and perturbed states of the bases in a base-pair.
                         Vector $\vec{Y}_n $ equals to the difference between $\vec{D}_n$ and $\vec{N}_n$, that is  $\vec{Y}_n = \vec{D}_n - \vec{N}_n$.}
	\label{fig:model}
\end{figure}

We shall utilize a one-dimensional lattice model of the DNA which accommodates these requirements,
in agreement with paper \cite{Georghiou}.
We follow the scheme introduced by El Hasan and Calladine, \cite{calladine}.
Aiming at a qualitative description of the DNA dynamics we  use a simplified set of variables, and
describe the relative position of the bases of a base-pair by  means of the vector $\vec Y$;
which is equal to zero when the base-pair is at equilibrium, see FIGs.\ref{fig:pairs}, \ref{fig:model}.
The relative positions of different
base-pairs are described  by  torsional angles $\varphi_n$  of the sugar-phosphate backbone.
They correspond to  deviations from the standard equilibrium twist of the double helix,
so that  a twist of the DNA molecule, which does not involve inter-strand motion or
mutual displacements of the bases inside the pairs, is determined by the torsional angles
$\varphi_n$  of rotation of the base-pairs about the axis of the double-helix.
Thus, at equilibrium,  a base-pair can be transformed into the next one by rotation through
the pitch angle $\Omega$ and simultaneous translation through a distance of neighbouring  base-pairs.
This picture of the conformation  of the DNA is in full agreement  with that of paper \cite{Georghiou}.

 We suppose that the size of DNA molecule is small enough that it can be visualized as a straight double helix, that is not larger than
the persistence length.  Hence the number of base-pairs, $ N \le 150 $, approximately. Consequently,
the  twist energy of the molecule is given by the equation
$$
     \sum_n\, \left[
		    \frac{I_n}{2} \, \dot{\varphi}_n^2
		    +  \displaystyle{\frac{\tau_n}{2a^2}} \,
		       (\varphi_{n+1} - \varphi_n)^2
		 \right]
$$
in which $I_n$ is the moment of inertia  of the $n$---th base-pair, and $\tau_n$ are the twist
coefficients, which may change from one base-pair to another. Interstrand motions should correspond to the relative motion,
or libration in terms of paper \cite{Georghiou}, of the bases inside the base-pairs, therefore the kinetic energy due
to this degree of freedom may be cast in the form
$$
  \sum_n\, \frac{M_n}{2}\, \dot{\vec Y}_n^2
$$
where $M_n$ is the effective mass of the $n$---th base-pair.

For each base-pair we have the reference frame in which z-axis corresponds to the axis of the double helix,
y-axis to the long axis of the base-pair, x-axis perpendicular to z- and y- axes (see Fig. 1 of paper \cite{calladine}).
At equilibrium the change in position of neighbouring base-pairs is determined only by the twist angle $\Omega$ of
the double helix.  We shall assume $\Omega = 2 \pi / 10$ as for the B-form of DNA.
To determine the energy due to the inter-strand displacements  we need to find the strain taking into account
the helical structure of our system. Aiming at a simplified picture of inter-strand motion, we
confine ourselves  to the torsional degrees of freedom of the double lattice and assume the vectors $\vec Y_n$ being
parallel to x-y plane, or two-dimensional. Consider the displacements $\vec Y_n,\, \vec Y_{n+1}$
determined within the frames of the two consecutive base-pairs, n, \, n+1. Since we must compare the two vectors in the same frame,
we shall rotate  the vector $\vec Y_{n+1}$ to the frame of the n-th base pair,
$$
  \vec Y^{\, back}_{n+1} =  R^{-1}(\varphi)\, \vec Y_{n+1}
$$
Here $R^{-1}(\varphi)$ is the inverse matrix of the rotation of the n-th frame to the (n+1)-one given by the equation
\begin{equation}
  R(\varphi
  ) = \left[
	      \begin{array}{ll}
		\cos \varphi  & - \sin \varphi   \\
		\sin \varphi
  &   \cos \varphi
	      \end{array}
	    \right] \label{rot}
\end{equation}
The matrix $R$ is 2 by 2 since the vectors $\vec Y_n$ are effectively two-dimensional. Then the  strain  caused by the displacements of the base-pairs
is determined by the difference
$$
   \vec Y^{\, back}_{n+1}  - \vec Y_n
$$

It is important that the angle $\varphi
$ is given by the twist angle, $\Omega$, describing the double helix, in conjunction with  the torsional angles $\varphi_n$, so  that
$$
  \varphi = \Omega + \varphi_{n+1} - \varphi_n
$$
Therefore, the energy due to the {\it inter-strand} stress reads
$$
    \sum_n \left\{
	       \frac{M_n}{2}\, \dot{\vec Y_n}^2
	       + \displaystyle{\frac{k_n}{2a^2}} \,
		  \left[ R^{-1}(\Omega + \varphi_{n+1} - \varphi
_n)\, \vec Y_{n+1}
		    - \vec Y_n
		  \right]^2
	    \right \}
$$
where $k_n$ are elastic constants determining the librational motion of base-pairs.
It is important that they may change from one base-pair to another.
The equilibrium position of the double helix is the twisted one determined by $\Omega$, with all
$\varphi_n$ being equal to zero. Combining the formulae given above we may write down the total energy
of the DNA molecule in the form
\begin{eqnarray}
  {\cal H} &=&
       \sum_n\, \left[
		    \frac{I}{2} \, \dot{\varphi}_n^2
		    +  \displaystyle{\frac{\tau_n}{2a^2}} \,
		       (\varphi
_{n+1} - \varphi
_n)^2
		 \right] \nonumber \\
       &+& \sum_n \left\{
	       \frac{M_n}{2}\, \dot{\vec Y_n}^2
	       + \displaystyle{\frac{k_n}{2a^2}} \,
		  \left[ R^{-1}(\Omega + \varphi_{n+1} - \varphi
_n)\, \vec Y_{n+1}
		    - \vec Y_n
		  \right]^2
	       + \frac{\epsilon_n}{2}\, \vec Y_n^2
	    \right \}    \label{main}
\end{eqnarray}
in which  $k_n$ and $a$ are the torsional elastic constants and the inter-pairs distance, correspondingly. In summations given above
n is the number of a site corresponding to the n-th base-pair, and $ n= 1,2, \ldots, N $, $N$ being the number of pairs in the segment of
the DNA under consideration. The last term, $\epsilon_n \vec Y^2/2$ accommodates  the energy of the inter-strand {\it separation}
due to the {\it slides of the bases inside the base-pairs}.

Thus, we have the model of a molecule of the DNA which may be visualized as $(1 + \epsilon )$---dimensional one, in the sense that it is one dimensional, from the formal point of view, and at the same time {\it accomodates librational motions of base-pairs} into ``two transversal dimensions'' outside the axis of the duplex. This is due to   $\vec Y_n$  being directed outside the axis of a molecule,   while scalar angles $\varphi_n$  describe the twist of the sugar-phosphate backbone. Our next step is to split the above $(1 + \epsilon)$---dimensional lattice into three interacting linear chains. To that end we shall cast our variables in a  complex form.
\begin{figure}[h]
	\center{\includegraphics[width=11cm]{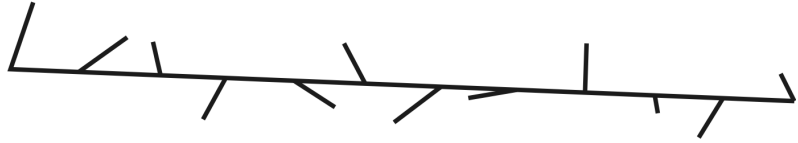}}
	\caption{Lattice of vectors $\vec Y_n$ of  relative positions of bases in a base-pair.}
	\label{fig:strand}
\end{figure}
Let us introduce complex quantities  $z_n$ in accord with the equations
$$
	\vec{Y}_n = (Y_n^1;Y_n^2) \rightarrow z_n = Y_n^1+iY_n^2
$$
Then we may cast the expression for the energy, $ E = T \; + \; U$,  in the form
$$
	T =  \sum_{n=0}^N \frac{I_n \dot{\varphi}_n^2}{2} + \sum_{n=0}^N \frac{m_n\dot{z}_n\dot{z}_n^*}{2} ,
$$
for the kinetic energy, and
$$
	U = \sum_{n=0}^N \frac{\epsilon_n z_n z_n^*}{2}
            \;+  \; \sum_{n=0}^{N-1} \frac{\tau_n}{2a^2}|\varphi_{n+1}-\varphi_n|^2
            \; + \; \sum_{n=0}^{N-1} \frac{k_n}{2a^2}
                                \left| z_{n+1} - \displaystyle{ e^{ \displaystyle{ i(\Omega \; + \; \varphi_{n+1}\; - \; \varphi_n) } } } z_n
                                \right|^2.
$$
for the potential one.  On using the substitution
$$
	z_n = e^{i \, n \cdot \Omega} \; y_n
$$
we may cast the above equations  in the form
$$
	T =  \sum_{n=0}^N \frac{I_n \dot{\varphi}_n^2}{2}
               \; + \; \sum_{n=0}^N \frac{m_n \dot{y}_n \dot{y}_n^*}{2} ,
$$
and
$$
	U = \sum_{n=0}^N \frac{\epsilon_n y_n y_n^*}{2}
            \; + \; \sum_{n=0}^{N-1} \frac{\tau_n}{2a^2}|\varphi_{n+1} \; - \; \varphi_n|^2
            \; + \; \sum_{n=0}^{N-1} \frac{k_n}{2a^2}
                             \left| y_{n+1} \; - \;  e^{ \displaystyle{ i(\varphi_{n+1} \, - \,\varphi_n) } } y_n \right|^2 .
$$
We cosider only small deviations from the equilibrium, and  assume that  difference $(\varphi_{n+1} - \varphi_n)$ is small. Therefore, we have the equation
$$
 e^{ { \displaystyle i(\varphi_{n+1} \; - \; \varphi_{n}) }} = 1 \; + \; i(\varphi_{n+1} \; - \; \varphi_n) \; + \; O((\varphi_{n+1} \; - \; \varphi_n)^2).
$$
In what follows we shall neglect terms of the second order in $\varphi$.

Within the above approximation  the Lagrangian equations of motion have the form
\begin{eqnarray}
         \label{eqn:phi}
	 a^2I_n \ddot{\varphi}_n  &=& \tau_{n-1} \varphi_{n-1} \; -\; (\tau_{n-1} \; + \; \tau_n)\varphi_n \; + \; \tau_n\varphi_{n+1}  \\
                               & & \; - \; k_n(a_nb_{n+1}  -  a_{n+1}b_n) \; + \; k_{n-1}(a_{n-1}b_n \; - \; a_nb_{n-1}) \; , \nonumber \\
	\label{eqn:a}
	 a^2 m_n \ddot{a}_n    &=& \; - \; a^2 \epsilon_n  a_n \; + \;  k_{n-1} a_{n-1} \; - \;  ( k_{n-1} \; + \; k_n ) a_n \; + \; k_n  a_{n+1}  \\
                               & & \; + \; k_n(\varphi_{n+1} \; - \; \varphi_n) b_{n+1}  \; - \; k_{n-1}(\varphi_n \; - \;\varphi_{n-1}) \, b_{n-1} \; , \nonumber \\
        \label{eqn:b}
         a^2 m_n \ddot{b}_n    &=& \; - \; a^2 \epsilon_n  b_n \; + \;  k_{n-1} b_{n-1} \; - \;  ( k_{n-1} \; + \; k_n ) b_n \; + \;  k_n  b_{n+1} \\
                               & & \; - \; k_n(\varphi_{n+1} \; - \; \varphi_n) a_{n+1} \; + \; k_{n-1} (\varphi_n \; -\; \varphi_{n-1}) a_{n-1} \; .\nonumber
\end{eqnarray}
$a_n$ and $b_n$ being the real and imaginary parts of $y_n$. Thus, we split the equations of motion into three parts, and
cast the system in a form of three interacting one-dimensional lattices of variables $a_n, \; b_n, \; \varphi_n$.
This mathematical device is crucial for the subsequent analysis of the dynamics.

It is worth noting that, in real life, a molecule of the DNA is not a perfect duplex, so that  constants $M_n,I_n,k_n,\epsilon_n$ are, generally,  random sequences.  Indeed, owing to different sets of  base-pairs, the conformational parameters of a molecule of the DNA, such as angles determining positions of base-pairs, suffer considerable, upto tens of per cent, deviations from the regular positions, \cite{calladine}. Further,  their being of the two types, adenine---thymine and guanine---cytosine, results in constants $M_n,I_n,k_n,\epsilon_n$   having different values for different pairs.   Therefore, equations (\ref{eqn:phi}) --- (\ref{eqn:b}) describe the motion of irregular interacting lattices. One can appreciate this circumstance according to ones lights. For one thing, it  results in formidable difficulties as far as the standard analytical treatment is concerned, for another, it is conducive to a phenomenon peculiar to molecules of the DNA in real life,
that is the  formation of localized excitations. In this paper we shall prefer the second option.

We feel that one can understand the phenomenon
of localized excitations within the framework of the theory for dynamic
excitations in random media worked out, years ago , by I.M.Lifshitz
and his colleagues,~\cite{Lif1} --- \cite{Lif5} (see
also the excellent review \cite{Maradudin}).  According to the
Lifshitz theory the dynamics of elastic excitations in random media,
for example  lattices of real crystals or amorphous systems, has two
important properties.  Firstly, there exists forbidden bands in the
spectrum of elastic excitations, such as phonons.  Secondly, there
are excitations localized in bounded regions of the medium,  their
characteristic frequencies being located at the edges of the
forbidden zones.  The key observation due to I.M. Lifshitz was that defects of media
could serve centers for the localization of elastic excitations, see
\cite{Lif1} --- \cite{Lif5}. In his first paper on the
subject he considered defects caused by an atom of isotope that has
a mass different from the regular one, and showed that the defect
could generate localized vibrations at its side. Later, the approach
was further extended to defects of a more general kind and higher
dimensions, see \cite{Maradudin}, for example plains of defects, and
also boundaries of crystals.

The study of spectra for disordered, or random, lattices has
followed mainly two paths.  A number of papers employed analytical
methods, which generally could provide only qualitative description
of the spectra, see \cite{Maradudin}.  It was P.Dean, \cite{Dean},
who developed a powerful computer technic for analysis situs of  the
frequency distribution of disordered lattices.
The results by P.Dean comply with the Lifshitz theory, and
provide a powerful insight into a specific
structure of spectra.


\section{ Hypersound spectroscopy.}
Phonon modes may result in the excessive absorption of hypersound in films and solutions of the DNA.
In fact, the passage of an acoustic wave may promote the transfer of molecules from the equilibrium state
to a state in which phonon modes of  molecules of the DNA are excited. The time delays in this process and
its reversal should lead to a relaxational dissipation of acoustic energy, and an  absorption of hypersound
at certain resonance frequencies corresponding to frequencies of the phonon modes. Equally important,
hypersound irradiation of molecules of the DNA could make for generatung  phonon modes.
Both the experimental data, \cite{Urabe} --- \cite{Edwards}, \cite{Lindsay},  and the theoretical arguments,
\cite{Mei}, \cite{Putnam}, \cite{Saxena_Zandt}, indicate that the frequencies of phonons of the DNA are in  GHz range,
and therefore one may expect resonance interaction between hypersound waves and phonons of the  DNA.
The availability of hypersound transducers, see \cite{Huynh}, \cite{Huynh2}, suggests that there are technical means
so as to employ high resolution  acoustical spectroscopy for studying phonon modes of the DNA in GHz range.

The alleged attenuation  requires a careful choice of right samples of the DNA for studying phonon modes.
By now films of the DNA are generally employed to that end. Perhaps, this could make for diminishing
the attenuation effects mentioned above. In fact, solutions of the DNA are not the best proposition
owing to  the attenuation effects  being  large  in this case. But, any way, it should be very interesting to use
liquid crystallin phases of the DNA, as was done for inelastic x-ray scattering, see paper \cite{Rupprecht2}
and references therein.

The attenuation of phonon modes is closely related to the problem of hydration of the DNA.
It is alleged that there are two relaxational processes in a hydrated molecule of the DNA
due to the primary and the second hydration shell with relaxational times $\tau_1  = 4 \times  10^{-11}$
and $\tau_2  = 2 \times 10^{-12}$ \, sec. The residence time for a molecule of water at grooves of
a molecule of the DNA is estimated as $0.2 \times 10^{-9} \div 0.4 \times 10^{-9} $ \, sec, \cite{Kumar}.
Phonon dynamics having characteristic times $10^{-9} \div 10^{-12}$ \, sec,
there is a need for using the generalized hydrodynamics in GHz range, \cite{Frenkel}, to estimate, for example,
the size of dissipation forces acting on a vibrating molecule. Presently, this problem is hard to solve, and we
have got only the information on the dispersion and attenuation of sound waves in this range, required
for the theoretical treatment of the Mandelstam---Brillouin light scattering.

But, curious enough, one may expect that the strong attenuation of some phonon modes could bring about the
generation of others because of the attenuation  due to the hydration shell mentioned above.
The latter primarily involve only  external regions of the duplex, or according to  paper \cite{Georghiou}
does not directly affect  the librations of base-pairs. Hence, we may suggest that an acoustic wave could drag the molecule,
promote its rotational motion, and  generate librations of base-pairs owing to the internal dynamics of the duplex.
We may infer from equations (\ref{eqn:phi}) --- (\ref{eqn:b}) that our model allows for the process. The problem is to
write down a reasonable force of molecular-liquid interaction.

For the convenience of numerical simulation it is worthwhile to choose appropriate scales for
length, mass, and time, which  agree with the conformational structure of the DNA.
We shall take the following quantities as the DNA units
\begin{itemize}
 \item $M = 10^{-22} \, gr$ as unit of mass, by the order of magnitude close to the mass of a base-pair;\\
 \item $L = 3 \times 10^{-8} \, cm$ as unit of length, close to the distance between neighbouring base-pairs; \\
 \item $T = 10^{-13} \, sec$ as unit of time; corresponding to the upper edge of phonon frequencies of the DNA.
\end{itemize}

It is important that the sound velocities, $c$, of a molecule of  the DNA are of  order $10^5 \, cm / sec$,
according to various experimental and theoretical estimates, \cite{Rupprecht}, \cite{Rupprecht2}.
In the units indicated above we have therefore $c \sim 0.3$.
The moment of inertia, $I_n$, of a base-pair is equal by orders of magnitude $m_n \, R^2$,
where $R$ is the radius of a molecule of the DNA, that is $\sim 10$ \AA.
Therefore, we have $I_n \sim 10$ in the DNA units introduced above.
To obtain numerical estimates for the constants $k_n,  \tau_n, \epsilon_n$ we shall require that
the values of sound velocities, $c$, and frequencies, $\omega$, of librations inside base-pairs, be of the orders of magnitude
$c = 0.3$ and $1$, respectively.
Since $k_n \, / \, m_n \sim c^2$ and  $\tau_n \, / \, I_n \sim c^2$, we have
$k_n \sim 0.1$ and $\tau \sim 1$, respectively.
For the libration frequency we have $\epsilon_n \, / \, m_n \sim 1$, if we assume that it be of order $1 \, GHz$.
For the viscosity coefficient of water we have, accordingly, $\mu \approx 0.3$.
Summarizing, we have the following characteristic quantities expressed in the DNA units
\begin{eqnarray*}
    c   & \sim &   0.3 \\
    I_n & \sim &   10  \\
    k_n & \sim &   0.1 \\
    \tau_n & \sim &  1 \\
    \epsilon & \sim &  1 \\
    \mu      & \sim &  0.3
 \end{eqnarray*}

As was mentioned above, the hydrodynamical theory, presently,  does not provide reliable theoretical
instruments for studying the interaction between a molecule and solvent,
and one should turn to a rule of thumb to find a solution. The interaction could be small enough, and
therefore proportional to the velocity of a molecule. Confining us to the picture given by paper \cite{Georghiou}, we may suggest
that the viscous force imposed on  the $n$-th cite of the molecule should have the form
\begin{equation}
      \label{eqn:ViscForce}
      d_n = - \, \gamma \dot{\varphi}_n
\end{equation}
where $\gamma$ is a dissipative constant. The size of $\gamma$ is difficult to assess. Estimates based on the conventional picture of liquid motion,
that is the Navier---Stokes one, are apparently far from reality because of small characteristic (``nano'') times, and the specific structure of water
close to the molecule (see \cite{Lindsay}), so that there is a problem as to whether it is reasonable to employ the usual hydrodynamic viscosity
coefficient in this situation. If we assume that it is possible, the dimension analysis gives for the moment of viscousity at site n
$$
  \frac{\gamma}{a^2} \propto \mu \, R^3
$$
or
$$
   \gamma \propto \mu \, a^2 \, R^3
$$
In the DNA units introduced above $ \mu \approx 0.3, \quad a = 1. , \quad R \approx 3 $
so that we have $\gamma \approx 10$.

In choosing  the drag due to the action of a sound wave on the molecule we have to take into account that
the wavelength of a hypersound wave in GHz region is by several orders of magnitude larger
than the molecular segment. The latter being of  a few hundred \AA,  the force depends only on time.
We assume also that the torsion of molecule due to the force is small, and  may take it in the form
\begin{equation}
      \label{eqn:Drag}
      f = A \, \varphi_n \, cos(\omega \, t)
\end{equation}
The main point is to determine the nature of the force and to assess its size.
If we confine ourselves to the {\it macroscopical} Navier---Stokes hydrodynamics, there are two  different pictures surfacing.

First, there could be an effect similar to that of the Rayleigh disc, \cite{Rayleigh}.  To see the point let us
neglect the dissipation and attribute the force to the streamlines of flow round a molecule.
There is the turning---moment $Q$  experienced by an oblate ellipsoid, or disk, in a vibrating fluid, \cite{Koenig}.
$$
  Q = \frac{4}{3} \, \rho \, R^3 \, v^2 \, \sin 2 \varphi
$$
The dimension analysis shows that there is the general formula for the turning---moment
$$
  Q \propto \rho L^3 \, v^2 \varphi
$$
where $L$ is the characteristic length of a body, $v$ velocity and $\rho$ the density of the fluid. In the case we are considering,
we may suggest that the rotation angle is small so as to employ
$\varphi$ instead of its sinus. Since the coefficient at $\ddot \varphi$ is the product $a^2 \, I_n$, we obtain the expression
$$
   A \propto \rho R^3 \, a^2 \, v^2
$$
Now we are in a position to assess the value of $A$ as regards the energy pumped in a sample. Let us recall that the density
$J_E$ of the energy current is given by the equation $J_E = \rho c v^2$, c being the velocity of sound in liquid.
We are interested only in rough estimates. Therefore we assume $c \sim 10^5 cm/sec, \; \rho \sim 1 \; gramme$.
For the energy current $J_E =  1 \; W/ cm^2$ we obtain $v  = 10 cm /sec$,  or $ v \sim 0.3, 10^{-4}$ in the DNA units.
In the equation  given above $R$ is the radius of the molecule, that is $\sim 10^{-7} \; cm$, or $R \sim 3$ in the DNA units.
Consequently, we have $A \sim 10^{-7}$. Compared with the values of $I_n, \; \gamma$,
it is too small to produce any appreciable effect.
In fact, even for $I_E \sim 1 \; kW/cm^2$ we obtain only  $A \sim 10^{-4}$, which is
also too small.  Therefore, the effect of the Rayleigh disk could not have any bearing on phonon modes of the DNA.
But it is worth noting that the above estimate is due to the use of the laws of {\it macroscopical hydrodynamics},
whereas there are no definite conclusions as to their validity on {\it microscopical scale}.

The second option is provided by the viscous interaction of the solvent with the molecule. We are looking for an analogue
of the familiar Stokes force " $6 \pi \mu R^2 v$ ``. Using the dimensional analysis we obtain the expression
\begin{equation}
  \label{eqn:ViscDragCoeff}
   A \propto \mu \, R^2 \, a^2 \, v
\end{equation}
For $J_E = 1 \; W / cm^2$ it provides the estimate $A \sim 10^{-4}$, or $A \sim 3 \cdot 10^{-3}$ for  $J_E = 1 \; kW / cm^2$.

The above estimates apparently preclude any opportunity for the observation of phonon modes of the DNA by means of hypersound pumping.
Similar arguments, \cite{Adair}, are generally put forward as regards  phonon modes studied by the sub-millimeter absorption spectroscopy.
But nonetheless the sub THz-phonon modes are observed, see \cite{Woolard} and the references therein. We feel that this discrepancy between
theory and experiment is likely to be due to the use of conventional hydrodynamics in the region where it does not work properly. In particular,
it results in the values of the coefficients $A$ and $\gamma$ in equations (\ref{eqn:phi})---(\ref{eqn:b}) that precludes the existence
of phonon modes. Therefore, in what follows we consider consequencies of  $A$ and $\gamma$ being outside the range
prescribed by the Navier---Stokes hydrodynamics, and choose appropriate values that permit the existence of phonon modes .

\section{Numerical simulation  of phonon modes.}
The equations of motion are obtained by inserting the  forces (\ref{eqn:ViscForce}), (\ref{eqn:Drag})
in equation(\ref{eqn:phi}). The  equations are nonlinear and require numerical simulation for their studying.
For the numerical simulation of the above equations we have used the algorithms of Verlet, LeapFrog, and the explicit and implicit Adams algorithms.
We have systematically made comparisons between results provided by different algorithms so as to escape possible mistakes and numerical artifacts.
We have taken the integration step $0.01$, or $0.001$ for verification, the unit of time being the period of external force generated by sound pumping..

We are going to simulate the generation of phonon modes by  pumping  hypersound.
According to  the equations of motion it is allowed owing to the term given by equation   (\ref{eqn:Drag}).
Thus, we shall generate firstly a $\varphi$-mode, and the latter shall promote the $a_n$- and the $b_n$-ones,
according to equations (\ref{eqn:a}) and (\ref{eqn:b}). As far as these modes are concerned,
we have a parametric excitation through the interaction terms in equations   (\ref{eqn:a}),(\ref{eqn:b}), and,
as is shown below, there may emerge a parametric resonance at a certain frequency $\omega_{ex}$ of the excitation pulse.
Here it should be noted that the usual theory of parametric resonance for the harmonic oscillator, \cite{Rayleigh},
cannot be directly employed because the equations (\ref{eqn:a}) --- (\ref{eqn:b}) of motion
for the interacting $\varphi_n, \, a_n, \, b_n$  chains are nonlinear, and we need to find the resonance frequencies
by ``trial and error'', Lifshitz's theory providing at this point  general recommendations.
\begin{figure}
    \center{\includegraphics[width=10cm]{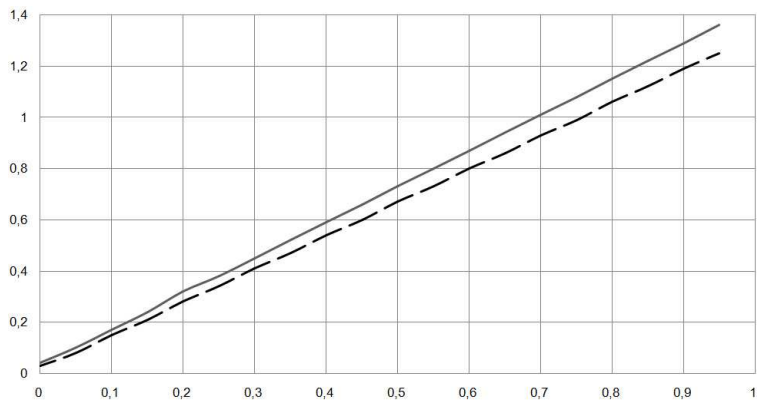}}
             \caption{Minimal  amplitude $A$ against dissipation $\gamma$.
                      Resonance  frequencies $1.3$ (dashed) and $1.4$ (solid).
	              The analytical fit is linear approximations   $ 1.296 \gamma + 0,019 $ and  $ 1.395 \gamma + 0,033 $,
		      respectfully.}
    \label{fig:dissipation}
\end{figure}

It is important that there exists the threshold  $A_{min}$ that depends on the dissipative coonstant $\gamma$,
the resonance taking place only  for amplitudes  of acoustic field which are large enough, $A \ge A_{min}$.

\begin{figure}
    \center{\includegraphics[width=10cm]{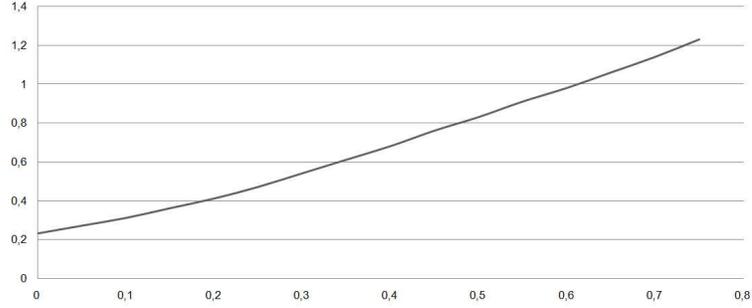}}
	   \caption{Threshold $A_{min}$ against the dissipation $\gamma$.
	            The excitation frequency equals 1.55 and deviates from  resonance one.
                    Number of sites $N = 300, \quad \mbox{mass} \quad m = 1$;
		    \quad elastic constants
		    $ \tau = 1 \quad \mbox{or} \quad 1.5 \quad (\mbox{equal probabilities});
		      \quad \epsilon = 1;
		      \quad k = 0.1 \quad \mbox{or} \quad 0.15
		      \quad (\mbox{equal probabilities});
		      \quad I = 10
		    $.
		    The value of $A_{min}$ at $\gamma = 0$ is not equal to zero.
		   }
    \label{fig:threshold}
\end{figure}

If the frequency of the acoustic wave is very close to the resonance one, $A_{min}$  depends linearly on
the dissipative constant $\gamma$, see FIG.\ref{fig:dissipation}, and for small $\gamma$, $\quad A_{min}$
is also small, so that there is no threshold value at $\gamma = 0$ and  $A_{min} = 0$.
If the excitation frequency appreciably deviates from the resonance one,
there is a threshold for $A_{min}$ at $\gamma = 0$, see FIG.\ref{fig:threshold}.
Thus, the numerical simulation of equations (\ref{eqn:phi}) --- (\ref{eqn:b}) is in agreement with the general theory of parametric resonance.

\begin{figure}
    \center{\includegraphics[width=12cm]{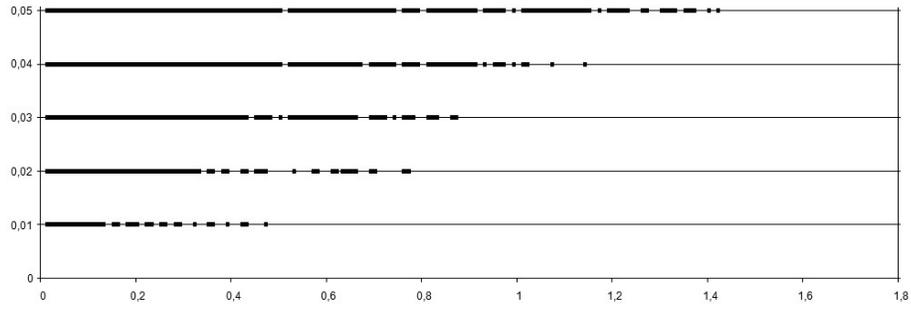}}
	    \caption{ Ranges of the resonance frequencies corresponding to the excitation amplitudes
	              $0.01, 0.02, 0.03, 0.04, 0.05 $  (thick horizontal segments);
	              $\epsilon = 1; k = 0.1 \quad \mbox{or} \quad  0.15$  (equal probabilities);
		      $I = 10$;
		      $\gamma = 0$.
		    }
           \label{fig:locarea}
\end{figure}

\begin{figure}
    \center{\includegraphics[width=12cm]{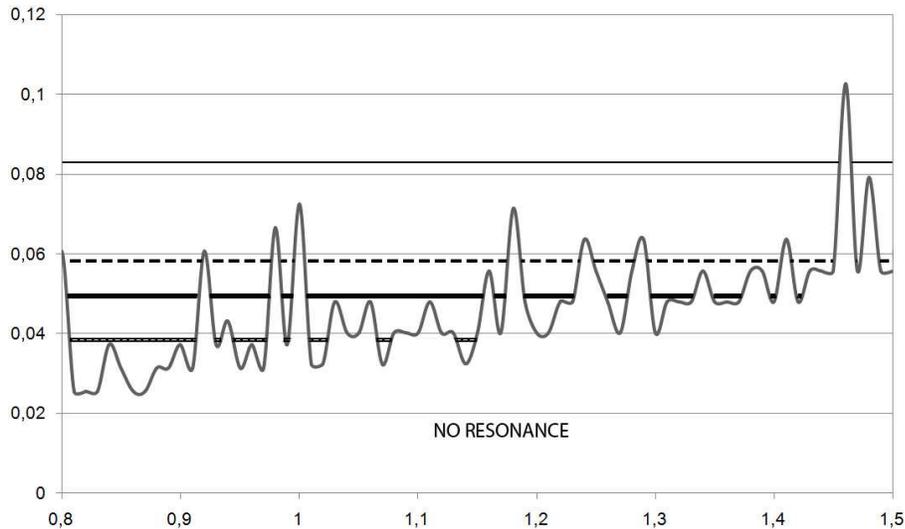}}
	    \caption{ Threshold curve $A_{min}(\omega)$ against excitation frequency (thin line).
	              Segments of the  frequencies $\gamma$ corresponding
		      to constant values $A \ge A_{min}(\omega)$ for which there is  resonance   (thick lines).
		    }
            \label{fig:explain}
\end{figure}

\begin{figure}
	\center{\includegraphics[width=12cm]{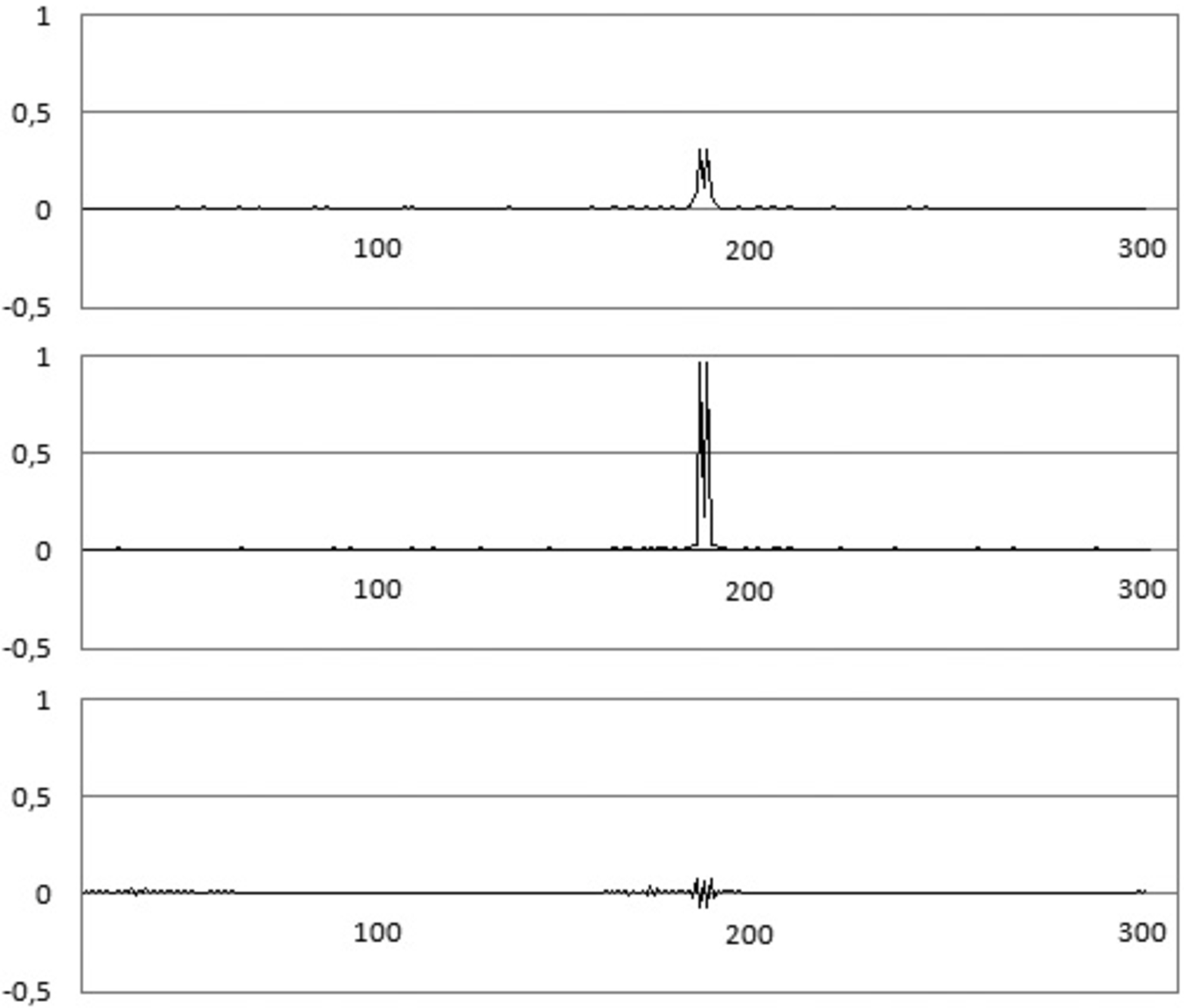}}
		\caption{ Dissipation $ \gamma = 0 $.   The DNA units are employed.
		          Localized excitation on a random lattice;
		          $ N=300,  m = 1; \tau = 1 \quad \mbox{or} \quad 1.5 $ (equal probabilities);
		          $ \epsilon  = 1; k = 0.1 \quad \mbox{or} \quad  0.15 $ (equal      probabilities);
			  $ I = 10 $;
			  amplitude $ A = 0.08;  $  excitation frequency $ \omega = 1.5 $.
			  Time elapsed 300 periods of the acoustic wave.
			  The  graphs correspond to the $a-, b-, \varphi-$modes (view from the top).
			}
	       \label{fig:localization}
\end{figure}

\begin{figure}
	\center{\includegraphics[width=12cm]{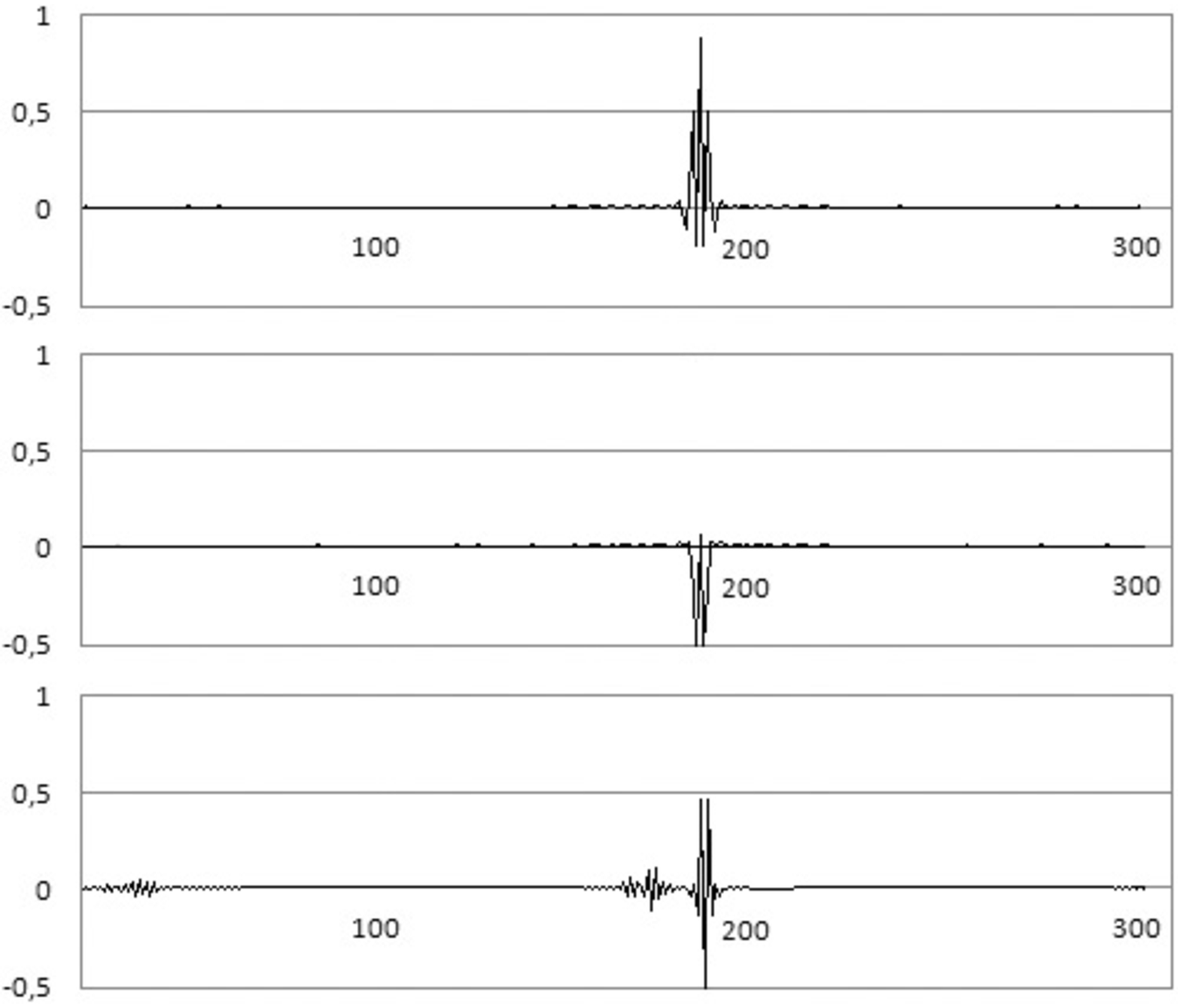}}
		\caption{ Dissipation $ \gamma = 0.01 $.   The DNA units are employed.
		          Localized excitation on a random lattice;
		          $ N=300,  m = 1; \tau = 1 \quad \mbox{or} \quad 1.5 $ (equal probabilities);
		          $ \epsilon  = 1; k = 0.1 \quad \mbox{or} \quad  0.15 $ (equal probabilities);
			  $ I = 10 $;
			  amplitude $ A = 0.1;  $  excitation frequency $ \omega = 1.489 $.
			  Time elapsed 1000 periods of the acoustic wave.
			  The  graphs correspond to the $a-, b-, \varphi-$modes (view from the top).
			}
	       \label{fig:loc01}
\end{figure}

\begin{figure}
	\center{\includegraphics[width=12cm]{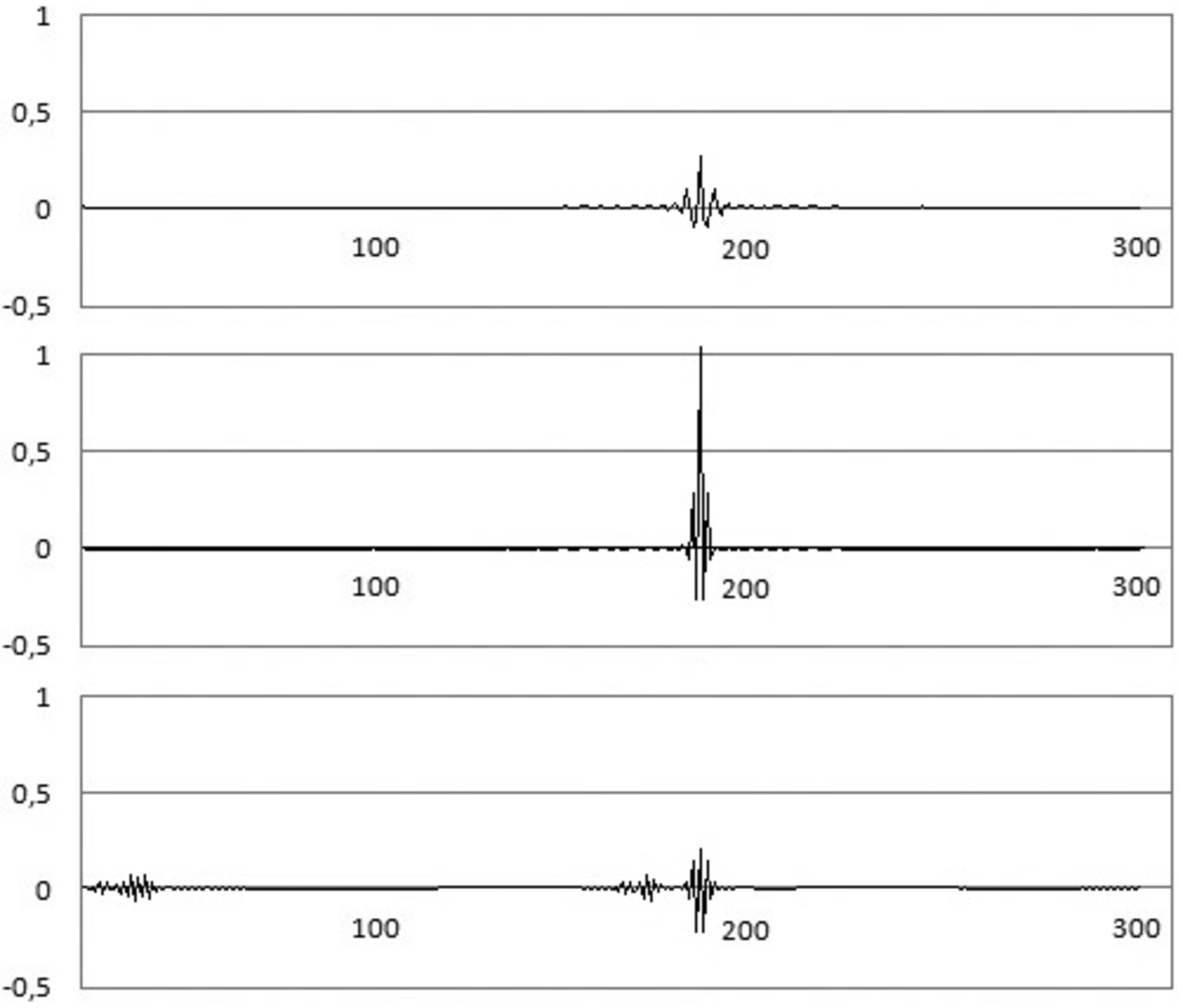}}
		\caption{ Dissipation $ \gamma = 0.02 $.   The DNA units are employed.
		          Localized excitation on a random lattice;
		          $ N=300,  m = 1; \tau = 1 \quad \mbox{or} \quad 1.5 $ (equal probabilities);
		          $ \epsilon  = 1; k = 0.1 \quad \mbox{or} \quad  0.15 $ (equal  probabilities);
			  $ I = 10 $;
			  amplitude $ A = 0.1;  $  excitation frequency $ \omega = 1.489 $.
			  Time elapsed 1300 periods of the acoustic wave.
			  The  graphs correspond to the $a-, b-, \varphi-$modes (view from the top).
			}
	       \label{fig:loc02}
\end{figure}

\begin{figure}
	\center{\includegraphics[width=12cm]{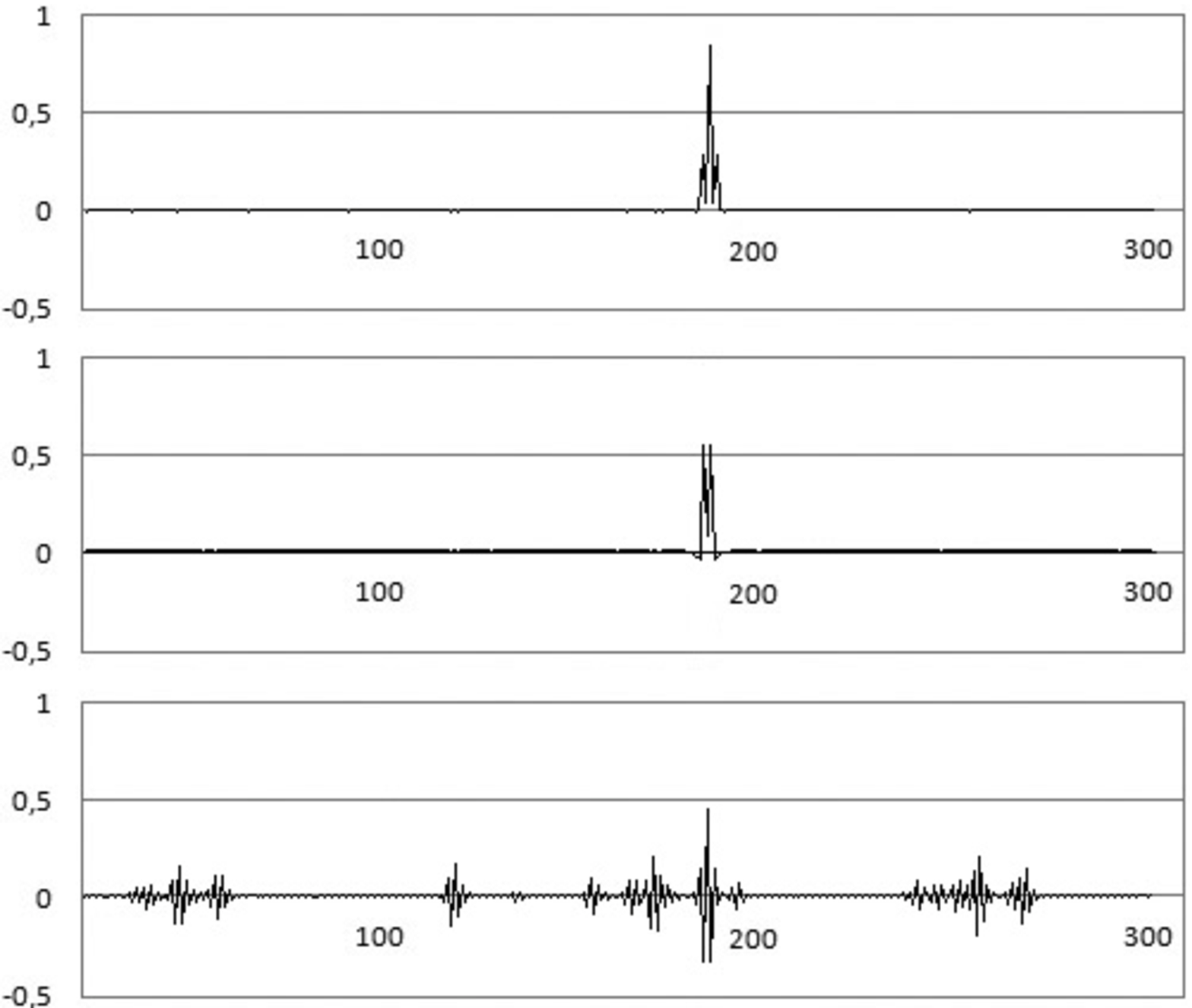}}
		\caption{ Dissipation $ \gamma = 1 $.   The DNA units are employed.
		          Localized excitation on a random lattice;
		          $ N=300,  m = 1; \tau = 1 \quad \mbox{or} \quad 1.5 $ (equal probabilities);
		          $ \epsilon  = 1; k = 0.1 \quad \mbox{or} \quad  0.15 $ (equal  probabilities);
			  $ I = 10 $;
			  amplitude $ A = 1.7;  $  excitation frequency $ \omega = 1.489 $.
			  Time elapsed 400 periods of the acoustic wave.
			  The  graphs correspond to the $a-, b-, \varphi-$modes (view from the top).
			}
	       \label{fig:loc02}
\end{figure}

\begin{figure}
	\center{\includegraphics[width=12cm]{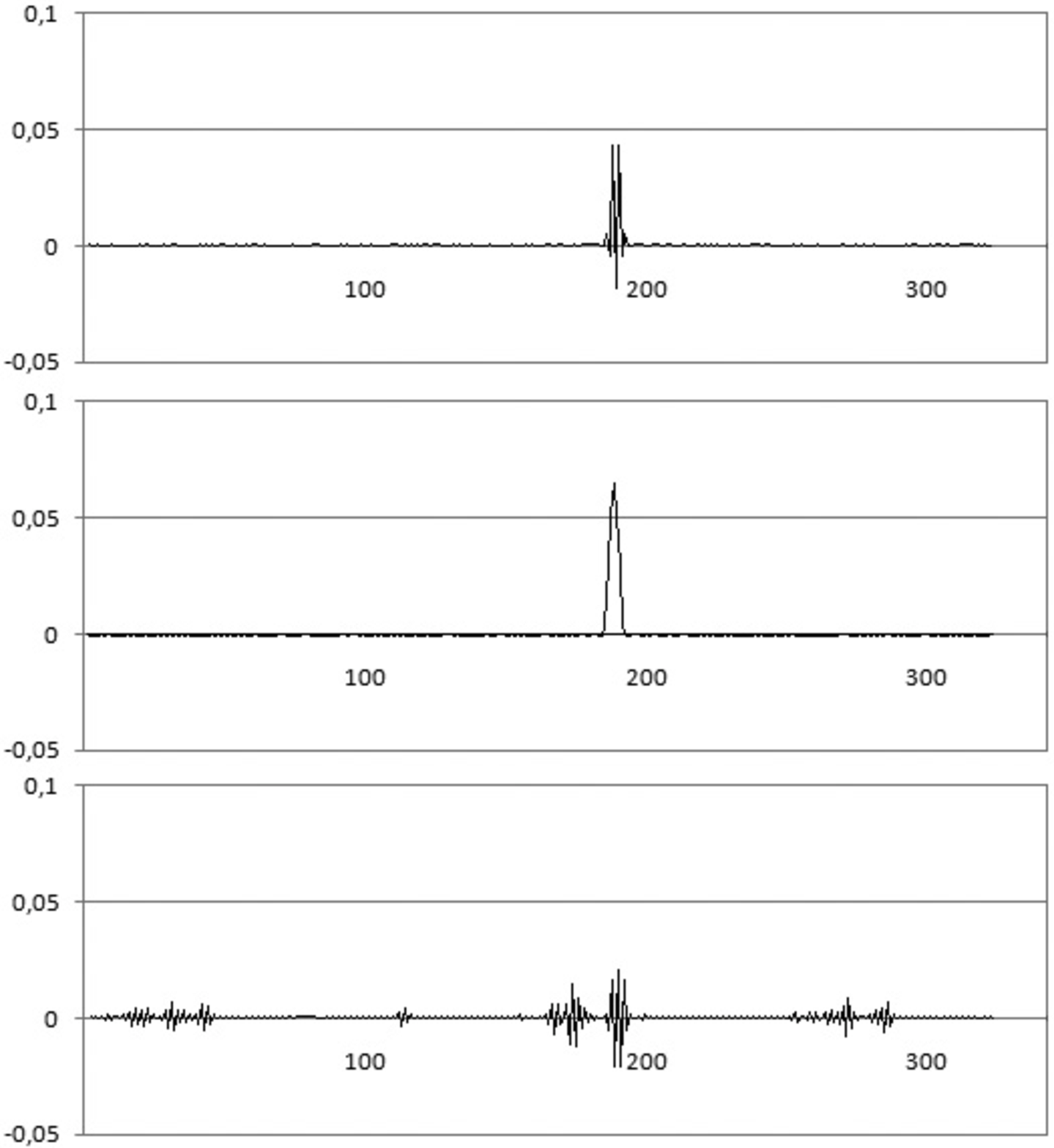}}
		\caption{ Dissipation $ \gamma = 0.8 $.   The DNA units are employed.
		          Localized excitation on a random lattice;
		          $ N=300,  m = 1; \tau = 1 \quad \mbox{or} \quad 1.5 $ (equal probabilities);
		          $ \epsilon  = 1; k = 0.1 \quad \mbox{or} \quad  0.15 $ (equal  probabilities);
			  $ I = 10 $;
			  amplitude $ A = 1.3;  $  excitation frequency $ \omega = 1.489 $.
			  Time elapsed 760 periods of the acoustic wave.
			  The  graphs correspond to the $a-, b-, \varphi-$modes (view from the top).
			}
	       \label{fig:loc08}
\end{figure}

Localization of a phonon mode excited by the resonant acoustic wave is illustrated
in FIG.\ref{fig:localization} for zero-dissipation case $\gamma = 0$,
in FIG.\ref{fig:loc01} for the dissipative constant $\gamma=0.01$, and in FIG.\ref{fig:loc02} for $\gamma=0.02$.
Our results thus in agreement with Van Zandt and Saxena,   \cite{Saxena}, who predicted the phonons in the submillimiter range
corresponding to  localized  excitations spread over several base pairs.

\section{Conclusion: a need for acoustic spectroscopy of the DNA}
We feel that the acoustic spectroscopy  could be helpful in analyzing the conformational structure of the DNA
through the analysis of its phonon modes.    The model we have constructed qualitatively agree with the experimental data
by providing the existence of the phonon modes and the reasonable orders of magnitude for their frequencies.
It implies  a phonon localization due to the random structure of the duplex, and
shows that it would be worthwhile to study the action of hypersound on molecules of the DNA.
It is important that absorption peaks for hypersound could be expected
at frequencies corresponding to the parametric resonance of phonon modes of the DNA under hypersound pumping.
In this respect the parametric resonance could be a valuable means for studying the vibrational modes of the DNA.

There still remains   a problem of accommodating the dissipative effects. The standard Navier---Stokes model promotes
a size of dissipation that exclude the existence of phonon modes, but they are observed in real life, \cite{Woolard}.
The circumstance challenges  theoreticians to working out an adeqate framework for the hydrodynamics at molecular scale.
The part played by the dissipation, perhaps different from the Navier---Stokes one,
could be crucial, and    torsional modes of the backbone could make for
understanding its nature on the nano-scale. Hypersound acoustic spectroscopy
could be instrumental in this respect. To that end liquid samples may turn out  to be more interesting,
solutions of the DNA  providing a probe into the physics of liquid state. The observation of
the parametric resonance we have indicated could serve a means for understanding the hydrodynamics on
molecular scale.

This paper was partially supported by the Russian Foundation for Fundamental Research, grant \#  09-02-00551-a.

\end{document}